\documentclass{amsart}
\usepackage{fullpage}
\usepackage{amsmath,amssymb} 
\usepackage[all]{xy}
\usepackage{mathbbol} 
\usepackage{bussproofs}

\SelectTips{cm}{}

\newcommand{\bZ}{\mathbb{Z}}
\newcommand{\bQ}{\mathbb{Q}}
\newcommand{\To}{\!\Rightarrow\!}

\newcommand{\catS}{{\mathbf{S}}} 
\newcommand{\catT}{{\mathbf{T}}} 
\newcommand{\catSp}{{\mathbf{S}}} 
\newcommand{\catTh}{{\mathbf{T}}} 
\newcommand{\spH}{\mathcal{H}} 
\newcommand{\spC}{\mathcal{C}} 
\newcommand{\For}{\mathrm{Form}} 
\newcommand{\Theo}{\mathrm{Theo}} 
\newcommand{\sk}{{\mathbf{E}}} 
\newcommand{\skSp}{{\mathbf{E}_{\catSp}}}
\newcommand{\skTh}{{\mathbf{E}_{\catTh}}}

\newcommand{\skThMP}{{\mathbf{E}_{\catTh,\MP}}}
\newcommand{\Hom}{\mathrm{Hom}}

\newcommand{\Rea}{\mathrm{Real}}
\newcommand{\Func}{\mathrm{Func}}
\newcommand{\Log}{\mathbf{Log}} 
\newcommand{\Set}{{\mathbf{Set}}}

\newcommand{\IM}{IM}
\newcommand{\MP}{MP}
\newcommand{\round}{\circlearrowright}
\newcommand{\ol}{\overline}
\newcommand{\Yon}{\mathcal{Y}}
\newcommand{\YonTh}{\mathcal{Y}_{\catTh}}
\newcommand{\YonSp}{\mathcal{Y}_{\catSp}}
\newcommand{\supseteqfull}{\supseteq_{\mathrm{full}}}
\newcommand{\monoto}{\xymatrix@C=1pc{\ar@{>->}[r] & \\ }}
\newcommand{\fracto}{\,\rightarrow\,} 

\newcommand{\mypar}[1]{\medskip\textbf{#1}} 

\begin{document}

\title{LOGICAL RULES AS FRACTIONS 
and LOGICS AS SKETCHES} 
\author{  Dominique Duval}
\address{Univ. Grenoble Alpes, CNRS, Grenoble INP, LJK, 38000 Grenoble, France}
\date{3 July 2018} 
\maketitle

\begin{itemize}
\item[]\textbf{Abstract.}
In this short paper, using category theory,
we argue that logical rules can be seen as fractions
and logics as limit sketches. 
\end{itemize}

\section*{INTRODUCTION}

This short paper relies on a talk given at the \emph{Universal Logic 2018}
conference, in the \emph{Category and Logic} workshop organised by
Peter Arndt.  
Quoted from the home page of \emph{Universal Logic}:
\begin{itemize}
\item[]
 \emph{
  ``Universal logic is a general theory of logical structures.
  Universal logic is not a new logic, it is a way of unifying the multiplicity
  of logics by developing general tools and concepts that can be applied
  to all logics.''
}
\end{itemize}

In this paper, using category theory,
we argue that logical rules can be seen as fractions
and logics as limit sketches, with the hope that these tools and concepts
can be applied to many kinds of logics.
A detailed presentation, with additional examples,
can be found in \cite{Du03,Du07,DD10,DD12}. 
The importance of categorical fractions for proofs and computations
was recognised independently in \cite{TW07}. 

For rules, it is a fact that
logical rules \emph{are written as} fractions $ \frac{H}{C} $,
with the conclusion as ``denominator'', 
and we argue that actually logical rules \emph{are} fractions $ \frac{C}{H} $,
with the hypothesis as denominator.
For a logic, first we define the theories (i.e., the  families of formulas 
which are closed under application of the rules) 
as the realisations of a sketch $\skTh$ where rules appear as arrows.
Then we derive from $\skTh$ a second sketch $\skSp$ and
a morphism $\sigma:\skSp\to\skTh$, such that
the specifications (i.e., all the families of formulas)
are the realisations of $\skSp$ and
the rules are fractions with respect to $\sigma$. 
An application to computational effects is mentioned at the end
of this paper, this subject is developed in \cite{DDR11,DDFR12,DDEP14,DDER14}. 

Here are some historical and recommended references for:
categories of fractions \cite{GZ67,Bo94},
sketches \cite{Eh68,BW95}
and locally presentable categories \cite{GU71,AR94}.
In this short paper we omit many technical issues, typically
issues related to size, choice, bicategories, etc.

\section*{I -- FRACTIONS}

\mypar{Categorical fractions.}

  Given two categories $\catS$, $\catT$ and a functor
  $ \xymatrix{\catS \ar[r]^{F} & \catT} $, 
  a \emph{fraction} $\dfrac{c}{h}:\spC \fracto\spH$ is
  (``essentially'') a cospan $(h,c)$ in $\catS$ (left) 
  such that $F(h)$ is invertible in $\catT$ (middle).
  We will use dashed arrows for representing ``both'' (right):
  
  $$ 
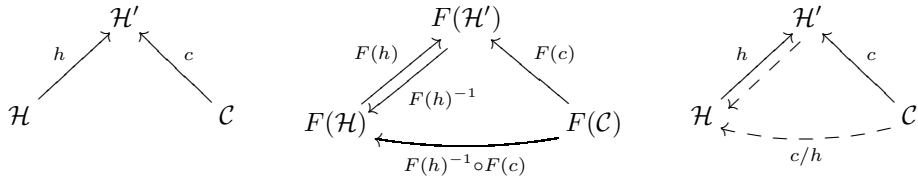
\xymatrix@C=2pc@R=2pc{
    & \spH' & \\
    \spH \ar[ru]^{h} &&
    \spC \ar[lu]_{c}  \\
  }  \qquad 
  \xymatrix@C=1.5pc@R=2pc{
    & F(\spH') \ar@<1ex>[ld]^{F(h)^{-1}} & \\
    F(\spH) \ar[ru]^{F(h)} &&
    F(\spC) \ar[lu]_{F(c)} \ar@/^/@<.5ex>[ll]^{F(h)^{-1}\circ F(c)} \\
  } \qquad
  \xymatrix@C=2pc@R=2pc{
    & \spH' \ar@{-->}@<1ex>[ld] & \\
    \spH \ar[ru]^{h} &&
    \spC \ar[lu]_{c} \ar@{-->}@/^/@<.5ex>[ll]^{c/h} \\
  }$$  

\mypar{Fractions, localisation, reflection} 

  A functor $ F:\catS \to \catT $ is: 
  \begin{itemize}
    \item a \emph{localisation} if
  it adds inverses for some morphisms in $S$; 
    \item a \emph{reflector} if 
  $\catT$ is a \emph{full subcategory} of $\catS$
  and $F$ is \emph{left adjoint} to inclusion:
    $$ \Hom_\catS(S,T) \cong \Hom_\catT(F(S),T)$$
  Then this adjunction is called a \emph{reflection} and this is denoted: 
    $$ \xymatrix@C=6pc{
        \catS \ar[r]_{F} \ar[r]^{\top} &
    \catT \ar@/_3ex/[l]_{\supseteqfull} \\
          }$$
  \end{itemize}

  \mypar{Theorem.} \cite{GZ67}.

  \emph{Every reflector is a localisation.}

\mypar{Example: the (usual) fraction $\frac{3}{4}$.}

On the integers (left), on the rationals (middle), and both (right): 
  $$
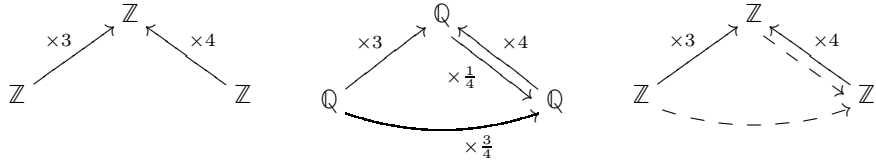
\xymatrix@C=2.5pc@R=1.5pc{
    & \bZ & \\
    \bZ \ar[ru]^{\times 3} && \bZ \ar[lu]_{\times 4} \\
}
\qquad
\xymatrix@C=2.5pc@R=1.5pc{
    & \bQ \ar@<-1ex>[rd]_(.4){\times \frac{1}{4}} & \\
    \bQ \ar[ru]^{\times 3}
      \ar@/_2ex/@<-.5ex>[rr]_(.65){\times \frac{3}{4}} &&
      \bQ \ar[lu]_{\times 4} \\
      }
    \qquad
    \xymatrix@C=2.5pc@R=1.5pc{
     & \bZ \ar@{-->}@<-1ex>[rd] & \\
    \bZ \ar[ru]^{\times 3} \ar@{-->}@/_2ex/@<-.5ex>[rr]
    &&
    \bZ \ar[lu]_{\times 4} \\
    }$$

    Thus, (usual) fractions are categorical fractions,
  with  $\catS=\textrm{Mod}(\bZ)$ the category of modules over the integers,
  $\catT=\textrm{Vect}(\bQ)$ the category of vector spaces over the rationals, 
  and $F:\textrm{Mod}(\bZ)\to\textrm{Vect}(\bQ)$ 
  the extension of scalars:
  $$ F(V) = \bQ \otimes V $$
Then $F(\bZ)=\bQ$ and
the integer 4 \emph{non-invertible} in $\bZ$ becomes 
the rational 4 \emph{invertible} in $\bQ$. 

\mypar{Logic, specifications, theories (informally).}

The following notions will be defined in the next sections. 

Given a \emph{logic}, with its \emph{formulas} and \emph{rules}, we say that:
  \begin{itemize}
  \item
    a \emph{specification} $S$ is a family of formulas;
  \item
    a \emph{theory} $T$ is a family of formulas 
    which is closed under application of the rules.  
  \end{itemize}

Let us assume the existence of: 
  \begin{itemize}
  \item a category $\catSp$ of \emph{specifications}
  \item a category $\catTh$ of \emph{theories}
  \item and a \emph{generating} functor 
    $F:\catSp\to\catTh$  
   such that $F(S)$ is the family of formulas (or \emph{theorems})
   deduced from the formulas (or \emph{axioms}) in $S$.
  \end{itemize}
  Then a logical rule is a categorical fraction wrt $F$.

\mypar{Example: the logical rule $\frac{p\;\; p\Rightarrow q}{q}$
  (\emph{Modus Ponens})}.

On specifications (left), on theories (middle), and both (right): 
  $$
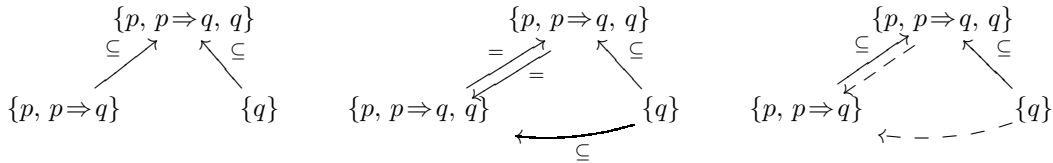
\xymatrix@C=-1pc@R=1.5pc{
    & \{p,\, p\To q,\, q\} & \\
     \{p,\, p\To q\}\; \ar[ru]^{\subseteq} && \{q\} \ar[lu]_{\subseteq} \\
} \qquad
\xymatrix@C=-.8pc@R=1.5pc{
    & \{p,\, p\To q,\, q\}
    \ar@<1ex>[ld]^(.4){=} & \\
    \{p,\, p\To q,\, q\}\quad
    \ar[ru]^{=} &&
    \{q\} \ar[lu]_{\subseteq}
      \ar@/^2ex/@<.5ex>[ll]^(.35){\subseteq} \\
} \qquad 
\xymatrix@C=-.5pc@R=1.5pc{
    & \{p,\, p\To q,\, q\}
    \ar@<.5ex>@{-->}[ld] & \\
    \{p,\, p\To q\}\;
    \ar@<.5ex>[ru]^{\subseteq} &&
    \{q\} \ar[lu]_{\subseteq} \ar@{-->}@/^2ex/@<.5ex>[ll] \\
  }$$

  Indeed, when \emph{modus ponens} is a rule of the logic,  
  let $S=\{p,\, p\To q\}$, then $F(S)= \{p,\, p\To q,\,q,\,...\}$:
  $S$ is a specification that \emph{does not contain} $q$
  while $F(S)$ is a theory that \emph{contains} $q$. 

\mypar{To sum up (I): Logical rules as fractions.}

More precisely, 
a logical rule $\dfrac{\spH}{\spC}$ is a fraction $\dfrac{c}{h}$:
``\emph{the hypothesis becomes invertible}''.

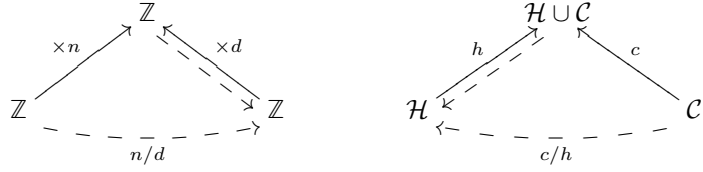
$$ 
\xymatrix@C=3pc@R=2pc{
    & \bZ \ar@{-->}@<-1ex>[rd] & \\
    \bZ \ar[ru]^{\times n} \ar@{-->}@/_/@<-1ex>[rr]_{n/d}  && 
    \bZ \ar[lu]_{\times d} \\
}
\qquad \qquad
      \xymatrix@C=2.5pc@R=2pc{
    & \spH \cup \spC \ar@{-->}@<1ex>[ld] & \\
    \spH \ar[ru]^{h} &&
    \spC \ar[lu]_{c} \ar@{-->}@/^/@<1ex>[ll]^{c/h} \\
      } $$
      
\section*{II -- SKETCHES}
      
 \mypar{Warning.}

 \emph{In this talk, \emph{sketch} always means \emph{limit sketch}.}

\mypar{Sketches and their realisations.}
 
 A \emph{sketch} $\sk$ is a \emph{presentation} for a category with limits $\ol\sk$. It is made of:
  \begin{itemize}
    \item  \emph{objects},
    \item ``\emph{morphisms}'' with only ``some'' identities and composition,
    \item and ``\emph{limits}'' with only ``some'' associated tuples, 
      \end{itemize}
  which become \emph{actual} objects, morphisms and limits in $\ol\sk$.
  We will use dotted arrows for denoting projections in limits. 

  A \emph{realisation} $R$ of a sketch $\sk$ is a \emph{set-valued model}
  of $\sk$: it maps each object, morphism and limit in $\sk$
to a set, function and limit in $\Set$.
Equivalently, a \emph{realisation} $R$ of $\sk$ is a limit-preserving functor
$R:\ol\sk\to\Set$.
Morphisms of realisations are ``natural transformations''  
and $\Rea(\sk)$ denotes the \emph{category of realisations} of~$\sk$. 

\mypar{The category $\Rea(\sk)$ is a kind of generalised presheaf.}

    \begin{itemize}
    \item A \emph{linear sketch} $\sk$ has only objects and morphisms
      (no limit);
      then $\Rea(\sk)=\Func(\ol\sk,\Set)$ is a \emph{presheaf category}.
      \smallskip
      \\ \textbf{Example.} 
      $\Rea(\xymatrix@C=3pc{ V & E \ar@<-.5ex>[l]_{s} \ar@<.5ex>[l]^{t} }) $
     is the category of directed graphs.
    \item In general, for a \emph{{[}limit{]} sketch} $\sk$, 
       $\Rea(\sk)$ is a \emph{locally presentable category}. 
      \smallskip
      \\ \textbf{Example.} 
       $ \Rea( \xymatrix@C=3pc{ M & M^2
    \ar@{.>}@<-.5ex>[l]_{s} \ar@{.>}@<.5ex>[l]^{t} \ar@/^2ex/@<1ex>[l]^{k} } )$ 
    is the category of magmas.  
    \end{itemize}

    \mypar{Remark.}

    \emph{``Many'' properties of presheaves 
    are still valid for locally presentable categories. }

\mypar{Logics as sketches.}
    
    We argue that it is possible to define \emph{a logic as a sketch}. 
    This will provide a very simple and very abstract algebraic 
    proposal for ``\emph{unifying the multiplicity of logics}'',
    or at least part of this multiplicity.

\mypar{Example: sketch for \emph{Modus Ponens}.} 

    As a basic example, starting from a logic 
    with \emph{Modus Ponens} $\Log_{\MP}$, 
    let us build the corresponding sketch $\skThMP$. 
    The logic $\Log_{\MP}$ is such that:
    \begin{itemize}
    \item The \emph{syntactic entities} are the
      \emph{formulas} ($\For$) and \emph{theorems} ($\Theo$),
      and each theorem is a formula.
      \item There are two rules: 
        the formation rule $ (\IM)$ states that
        \emph{if $p$ and $q$ are formulas then $p\To q$ is a formula}
        while the deduction rule $ (\MP)$ ensures that
        \emph{if $p$ and $p\To q$ are theorems then $q$ is a theorem}. 
        $$ (\IM) \quad \frac{p,\;q:\For}{p\To q\,:\For}
        \qquad \qquad
        (\MP) \quad \frac{{[}\;p,\;q,\;p\To q:\For\;{]} \quad p,\;p\To q\,:\Theo}{q:\Theo} $$
    \end{itemize}
    
A sketch  $\skThMP$ is now built in three steps.

        \begin{itemize}
          
        \item First, here is a sketch for the syntactic entities
          (where the arrow $\monoto $
          stands for a monomorphism, which is a kind of limit).
            A realisation $R$ of this sketch is made of
  a set of \emph{formulas} $R(\For)$ and a set of \emph{theorems} $R(\Theo)$,
  with $R(\Theo) \subseteq R(\For)$.
  $$
  \begin{array}{|l|}
    \hline
  \xymatrix{
    \For && \;\Theo \ar@{>->}[ll] \\ 
    } \\
    \hline
    \end{array}
  $$

  \item 
  Then, here is a sketch for the formation rule $(\IM)$,
  where the limits mean that $C_{\IM} = \For$
  and $H_{\IM} = \For^2$. 
  A realisation $R$ of this sketch is made of
  a set of formulas $R(\For)$,
  the sets $R(C_{\IM})=R(\For)$ and $R(H_{\IM})=R(\For)^2$,
 and a function $R(c_{\IM}):R(H_{\IM})\to R(C_{\IM})$
 that will be denoted $\; c_{\IM}(p,q) = p\To q$.
    $$ 
  \begin{array}{|l|}
    \hline
  \xymatrix@C=1pc@R=3pc{
    H_{\IM} \ar@{.>}@/_/[dr]
    \ar@{.>}@/^/[dr] \ar@<-.3ex>[rr]^{c_{\IM}} &&
      C_{\IM} \ar@{.>}[dl] \\ 
    & \For & \\ 
  }\\
    \hline
    \end{array}
  $$

\item
  And finally here is (a simplified version of) the sketch $\skThMP$, 
  where the limits mean that $C_{\MP}=\Theo$ and
  that $H_{\MP}$ is ``essentially'' $\Theo^2$.
  Drawing the precise limit diagram for $H_{\MP}$ is left as an exercice.
  It must be such that $R(H_{\MP})$ is the set of triples
  $(p,q,r)$ of formulas, with $p$ and $r$ theorems and with $r=p\To q$. 
  Thus, a realisation of $\skThMP$ is a \emph{theory}
  for the logic $\Log_{\MP}$: 
  $ \Rea(\skThMP) = \catTh_{\MP} $.
   $$ \skThMP = 
  \begin{array}{|l|}
   \hline
  \xymatrix@C=1pc@R=2.8pc{
    &&&& H_{\MP} 
         \ar@{.>}[dllll]
         \ar@{.>}[dll]
         \ar@{.>}@/_/[ddr]
         \ar@{.>}@/^/[ddr]
         \ar@<-.3ex>[rr]^{c_{\MP}} & &
         C_{\MP} \ar@{.>}[ddl] \\
    H_{\IM} \ar@{.>}@/_/[dr]
    \ar@{.>}@/^/[dr] \ar@<-.3ex>[rr]^{c_{\IM}} & &  
      C_{\IM} \ar@{.>}[dl] &&& \\ 
    & \For &&&& \;\Theo \ar@{>->}[llll] & \\ 
  } \\
    \hline
    \end{array}
  $$
  
\end{itemize} 

\mypar{To sum up (II): Logical theories as realisations of a sketch.}

If we define a \emph{logic} as a sketch $\skTh$,
then the category of \emph{theories} is the category of realisations
$\catTh=\Rea(\skTh)$.

At this point, we might define 
a \emph{model} of a theory $T$ in a theory $D$ as 
an arrow $M:T\to D$ in $\catTh$
and a \emph{rule} as an arrow $c:H\to C $ in $\skTh$.
However, this point of view is far from satisfactory,
mainly because there is no notion of \emph{specification}.
This is solved in Part (III), where in addition we recover
the fact that rules are \emph{fractions}, as in Part (I).

     \section*{III -- SKETCHES and FRACTIONS}

\mypar{Morphisms of sketches.}

A morphism of sketches is a generalised functor:
it maps objects, morphisms and limits
to objects, morphisms and limits.
Each morphism of sketches $\sigma : \sk_1 \to \sk_2 $ induces a functor
$ G : \Rea(\sk_2) \to \Rea(\sk_1)$
by mapping each realisation $R_2$ of $\sk_2$
to the realisation $R_2 \circ \sigma$ of $\sk_1$.             

\mypar{Theorem.} \cite{Eh68}.

\emph{The functor $G$ associated to $\sigma$ has a left adjoint. }
          
$$ \fbox{
          $ \xymatrix@C=6pc{
    \Rea(\sk_1) \ar@/_1.5ex/[r]_{F} \ar@{}[r]|{\top} &
    \Rea(\sk_2) \ar@/_1.5ex/[l]_{G}  \\
          }$
 } $$
 This means that each realisation of $\sk_1$ \emph{generates}
 a realisation of $\sk_2$.

\mypar{Cycles.}
 
  A ``\emph{cycle}'' in a sketch $\sk$ is defined by considering that
  projections are oriented both sides.

  \mypar{Example.}

  There is a cycle in the sketch for the formation rule $\;(\IM)\;\dfrac{p,\;q:\For}{p\To q\,:\For}$.
  
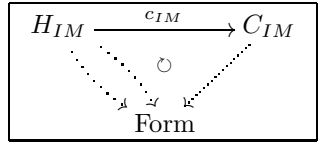
  $$ \begin{array}{|l|}
    \hline
  \xymatrix@C=1pc@R=2pc{
    H_{\IM} \ar@{.>}@/_/[dr]
    \ar@{.>}@/^/[dr] \ar@<-.3ex>[rr]^{c_{\IM}} &  \ar@{}[d]|(.4){\round} &
      C_{\IM} \ar@{.>}[dl] \\ 
    & \For & \\ 
  } \\ 
    \hline
    \end{array}
  $$
  \begin{itemize}
  \item Note:
  because of cycle ``$\round $'', in a \emph{theory} $T$ 
  for \emph{ALL} pairs of formulas $(p,q)$ there is a formula $p\To q$.
  \item    Required: in a \emph{specification} $S$ 
  for \emph{SOME} chosen pairs of formulas $(p,q)$ there is a formula $p\To q$.
  \end{itemize}

  \mypar{Breaking cycles.}

  \mypar{Theorem.} \cite{Du03}.
  
  \emph{ Cycles in a sketch can be broken ``in a reasonable way''.}

  The key point is to make some arrows \emph{partial}:
  
  \begin{center}
    replace \;\fbox{$
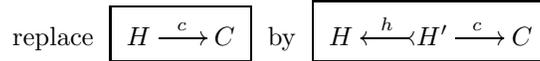
\xymatrix@C=1.5pc{H\ar[r]^{c} & C \\}$}\; by
    \;\fbox{$\xymatrix@C=1.5pc{ H & H'\ar@{>->}[l]_{h} \ar[r]^{c} & C }$}
    \end{center}
  By breaking the cycles in $\skTh$ we get a sketch $\skSp$
 and a morphism called a \emph{localiser} 
  $$ \fbox{$
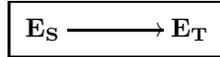
\xymatrix@C=3pc{ \skSp \ar[r] & \skTh }$} $$
 such that the corresponding adjunction is a \emph{reflection}.
 
  \begin{center}
        \fbox{
          $ 
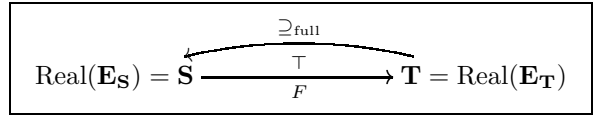
\xymatrix@C=6pc{
        \Rea(\skSp) = \catSp \ar[r]_{F} \ar[r]^{\top} &
    \catTh = \Rea(\skTh)  \ar@/_3ex/[l]_{\supseteqfull} \\
          }$
        }
        \end{center}

        \mypar{Definitions.}

        \emph{
A \emph{diagrammatic logic} is a sketch $\skTh$.
\\ 
   By breaking the cycles in $\skTh$ one gets a localiser
  $\sigma:\skSp \to \skTh$, thus a reflector $F:\catSp\to\catTh$.
 \medskip
    \begin{itemize}
    \item the category of \emph{theories} is $\catTh=\Rea(\skTh)$,
    \item the category of \emph{specifications} is $\catSp=\Rea(\skSp)$,
    \item the theory \emph{generated} by a specification $S$ is $F(S)$,
  \item a \emph{model} of a specification $S$ in a theory $D$ is 
    an arrow $M:S\to D$ in $\catSp$
    \\ {[} or equivalently, an arrow $M:F(S)\to D$ in $\catTh$ {]},
    \item a \emph{rule} is a fraction in $\skSp$ wrt $\sigma$. 
    \end{itemize}
  }
  
    These definitions can be illustrated as follows, 
using the \emph{Yoneda contravariant embedding}  
$\Yon : \sk^{op} \to \Rea(\sk)$, such that $\Yon(X) = \Hom_{\,\ol{\sk}\,}(X,-)$. 

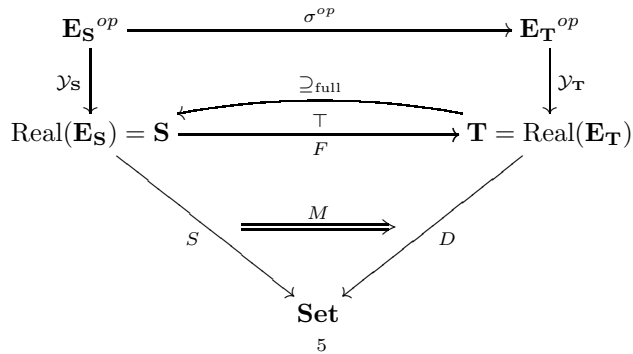
  $$ \xymatrix@C=1.5pc@R=2pc{
    \skSp^{op} \ar[rrrr]^{\sigma^{op}} \ar[d]_{\YonSp} &&&&
    \skTh^{op} \ar[d]^{\YonTh} \\ 
    \Rea(\skSp) = \catSp \ar[rrrr]_{F} \ar[rrrr]^{\top} \ar[rrdd]_{S} &&&& 
    \catTh = \Rea(\skTh)  \ar@/_3ex/[llll]_{\supseteqfull}
    \ar[lldd]^{D} \\
    & \ar@{=>}[rr]^{M} && &\\
    && \Set && \\
    }$$
    
Note that, thanks to $\Yon$, a \emph{rule} can also be seen as a fraction
in $\catSp$ wrt $F$ which is in the image of $\Yon$.
Then a \emph{proof} is any fraction in $\catSp$ wrt $F$,
and the density property of $\Yon$ (as expressed below)
ensures that proofs are built from rules. 

     \mypar{About the Yoneda contravariant embedding.}
     
  The embedding $\Yon : \sk^{op} \to \Rea(\sk)$ is ``nearly as nice''
  for \emph{locally presentable} categories as for \emph{presheaves}:
  \begin{itemize}
  \item $\Yon$ is \emph{faithful}, 
  \item $\Yon$ maps \emph{limits} to \emph{colimits}, 
  \item $\Yon(\sk^{op})$ is \emph{dense} in $\Rea(\sk) $:
    each realisation of $\sk$ is the \emph{colimit} of realisations
    in $\Yon(\sk^{op})$.
  \end{itemize}

  The category $\Rea(\sk)$
has all \emph{colimits} (like \emph{presheaves}) 
but they cannot be computed sortwise (unlike \emph{presheaves}).
This last property can be read as negative:
``\emph{computing colimits is not easy}''
or as positive: ``\emph{a large amount of theorems can be derived from
  a small amount of axioms}''.

\mypar{Example: breaking the cycle for rule (\IM).}

First in sketches: adding a rule is a morphism:
  $$  
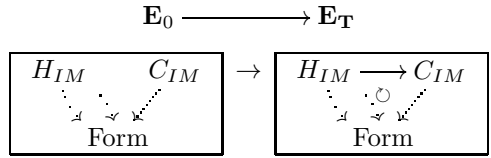
\xymatrix@C=4pc@R=2pc{
    \sk_0 \ar[r] &
    \skTh \\
}  $$
  $$   \begin{array}{|l|l|l|}
    \cline{1-1}\cline{3-3}
  \xymatrix@C=-.5pc@R=1pc{
    H_{\IM} \ar@{.>}@/_/[dr]
    \ar@{.>}@/^/[dr] &&
      C_{\IM} \ar@{.>}[dl] \\ 
    & \For & \\ 
  }   & \to
  & 
  \xymatrix@C=-.5pc@R=1pc{
    H_{\IM} \ar@{.>}@/_/[dr] \ar[rr]
    \ar@{.>}@/^/[dr] & \ar@{}[d]|(.35){\round} &
      C_{\IM} \ar@{.>}[dl] \\ 
    & \For & \\ 
  }
  \\
    \cline{1-1}\cline{3-3}
    \end{array}
  $$
  that gets factorised by breaking cycles:
  $$  
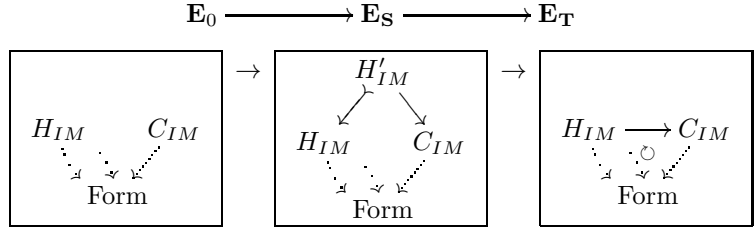
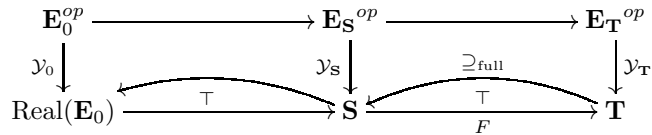
\xymatrix@C=4pc@R=2pc{
    \sk_0 \ar[r] &
    \skSp \ar[r] &
    \skTh \\
}  $$
  $$   \begin{array}{|l|l|l|l|l|}
    \cline{1-1}\cline{3-3}\cline{5-5} 
  \xymatrix@C=-.5pc@R=1pc{
    & \quad & \\
    H_{\IM} \ar@{.>}@/_/[dr]
    \ar@{.>}@/^/[dr] &&
      C_{\IM} \ar@{.>}[dl] \\ 
    & \For & \\ 
  } & \to
  & 
  \xymatrix@C=-.5pc@R=1pc{
    & H'_{\IM} \ar@{>->}[dl] \ar[dr] & \\  
    H_{\IM} \ar@{.>}@/_/[dr]
    \ar@{.>}@/^/[dr] &&
      C_{\IM} \ar@{.>}[dl] \\ 
    & \For & \\ 
  }   & \to
  & 
  \xymatrix@C=-.5pc@R=1pc{
    & \quad & \\
    H_{\IM} \ar@{.>}@/_/[dr] \ar[rr]
    \ar@{.>}@/^/[dr] & \ar@{}[d]|(.35){\round} &
      C_{\IM} \ar@{.>}[dl] \\ 
    & \For & \\ 
  }
  \\
    \cline{1-1}\cline{3-3}\cline{5-5} 
    \end{array}
  $$

  Now in realisations:
  
  $$  \xymatrix@C=6pc@R=1.5pc{
    \sk_0^{op} \ar[r] \ar[d]_{\Yon_0} &
    \skSp^{op} \ar[r] \ar[d]_{\YonSp} &
    \skTh^{op} \ar[d]^{\YonTh} \\ 
    \Rea(\sk_0) \ar[r]^{\top} &
    \catSp \ar[r]_{F}^{\top} \ar@/_3ex/[l] & 
    \catTh \ar@/_3ex/[l]_{\supseteqfull} \\
  } $$
Thus, focusing on $\Yon(-)(\For)$:
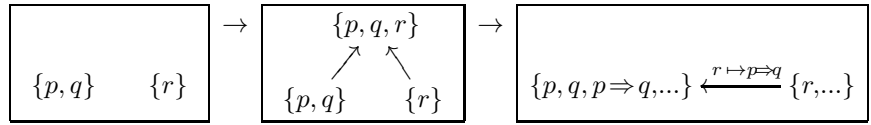
  $$   \begin{array}{|l|l|l|l|l|}
    \cline{1-1}\cline{3-3}\cline{5-5}
  \xymatrix@C=-.5pc@R=1pc{
    && \\ \{p,q\} & \qquad & \{r\} \\ } 
  & \to
  &  \xymatrix@C=-1pc@R=1pc{
    & \{p,q,r\} & \\ \{p,q\} \ar[ru] & & \{r\} \ar[lu] \\ } 
  & \to
  & \xymatrix@C=2.5pc@R=1pc{
    &  \\ \!\!\{p,q,p\To q,\!...\} & \{r,\!...\}\!\! \ar[l]_(.35){r \,\mapsto p\To q }  \\ } 
  \\
    \cline{1-1}\cline{3-3}\cline{5-5} 
    \end{array}
  $$
we get the \emph{fraction}:
   $$
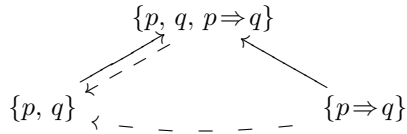
\xymatrix@C=1pc@R=1.5pc{
    & \{p,\, q,\, p\To q\}
    \ar@<.5ex>@{-->}[ld] & \\
    \{p,\, q\}\;
    \ar@<.5ex>[ru] &&
    \{p\To q\} \ar[lu] \ar@{-->}@/^1.5ex/@<.5ex>[ll] \\
  }$$

  \mypar{Morphisms of theories are presented by fractions of specifications.}
    
  \begin{itemize}
    \item 
      A \emph{specification} $S$ in $\catSp$ is a \emph{presentation}
      for the theory $T=F(S)$ in $\catTh$
    \item 
      A \emph{morphism} $s:S\to S'$ in $\catSp$ is a \emph{presentation} 
      for the morphism $F(s):F(S)\to F(S')$ in $\catTh$.
      \\ \smallskip
  In this way one gets \emph{SOME} morphisms $t:F(S)\to F(S')$ in $\catTh$.
    \\ \smallskip \quad \textbf{Example.} Every ring is a monoid.
\item
  A \emph{fraction} $ \dfrac{c}{h}: S \fracto S'$ wrt $F$
      is a \emph{presentation} for the morphism  
  $ F(h)^{-1}\circ F(c) :F(S)\to F(S')$ in~$\catTh$. 
      \\ \smallskip 
  In this way one gets \emph{ALL} morphisms $t:F(S)\to F(S')$ in $\catTh$. 
 \\ \smallskip \quad \textbf{Example.} Every boolean algebra is a ring.
  \end{itemize}

  \mypar{Finiteness issues.}

  It is a fact that every book, program, proof,... is \emph{finite},
  but logical theories are usually \emph{infinite}.
  \\  
  Let us say that a realization $R$ of a finite sketch $\sk$
  is \emph{finite} if the set $R(X)$ is finite for each $X$ in $\sk$.  
  \\  
  For a diagrammatic logic, when the sketch $\skTh$ is finite then: 
  \begin{itemize}
    \item the sketch $\skSp$ is finite, 
    \item the realisation $\Yon(X)$ is finite for each $X$ in $\skSp$, 
    \item and the hypothesis and conclusion of each rule are finite
  specifications. 
  \end{itemize}

   \mypar{To sum up (III): Logics as sketches and rules as fractions.}

   A \emph{diagrammatic logic} is a sketch, and 
   by breaking the cycles in this sketch one gets a localiser
   (between sketches), thus a reflector (between categories of realisations). 
   This provides a simple and abstract framework for defining the notions
   of theories, specifications, models, and rules as fractions. 
   Then \emph{morphisms} of diagrammatic logics are ``of course''
   defined as fractions of sketches. 

   \section*{IV -- Application: COMPUTATIONAL EFFECTS}

   The definition of a diagrammatic logic has been motivated by the study
   of imperative and object-oriented features in computer languages.
   Such features, called \emph{computational effects},
   can be seen from various points of view, corresponding to various logics
   related by non-trivial morphisms.
   We have built logics for reasoning about such programs without departing
   from their imperative or object-oriented flavour, with
   implementations in the Coq proof-assistant.
   Here is a toy example of this application. 
   
\mypar{The state effect in object-oriented programming.}

Let us consider the following piece of C++ code, for dealing with
toy bank accounts:

 \qquad \fbox{$\begin{array}{l}
  \texttt{Class BankAccount } \{... \\  
  \qquad  \texttt{int balance (void) const ;} \\
  \qquad  \texttt{void deposit (int) ;}\\
   ... \}\\
\end{array}$}

 Our goal is to associate to this piece of code a ``quasi-equational'' 
 specification. Here are three proposals. 

  \begin{itemize}

  \item The \emph{apparent} specification:
\\ $\qquad$ 
  \fbox{$\begin{array}{l}
  \texttt{balance}:\texttt{void}\to\texttt{int} \\
   \texttt{deposit}:\texttt{int}\to\texttt{void}  \\
 \end{array}$ }
\\ Here the object-oriented flavour is preserved 
BUT the intended interpretation is \emph{not} a model. 

\item The \emph{explicit} specification: 
\\ $\qquad$ 
  \fbox{  $\begin{array}{l}
  \texttt{balance}:\texttt{state}\to\texttt{int} \\ 
   \texttt{deposit}:\texttt{int}\times\texttt{state}\to\texttt{state} \\
 \end{array} $ }
\\ Here the intended interpretation is a model 
BUT the object-oriented flavour is \emph{not} preserved.  

\item \emph{decorated} specification: 
\\ $\qquad$  
  \fbox{  $\begin{array}{l}
  \texttt{balance}^\texttt{a}:\texttt{void}\to\texttt{int} \cr
  \texttt{deposit}^\texttt{m}:\texttt{int}\to\texttt{void} 
\end{array}$ }
\\ where the \emph{decorations} (superscripts) are:
\begin{itemize}
\item[] \texttt{m} for \emph{modifiers} (methods)
\item[] \texttt{a} for \emph{accessors}
  (``{\small\texttt{const}}'' methods)
\end{itemize}
Here the intended interpretation is a model
AND the object-oriented flavour is preserved. 

  \end{itemize}

  These three specifications live in three different diagrammatic logics,
  related by morphisms:
  a morphism from the decorated logic to the apparent logic,
  that forgets the decorations,
  and a morphism from the decorated logic to the explicit logic,
  that expands the code so as to make the semantics explicit.
  Our proofs lie in the decorated logic. 

  $$ \begin{array}{ccc}
\cline{2-2}
  &   
  \multicolumn{1}{|c|}{
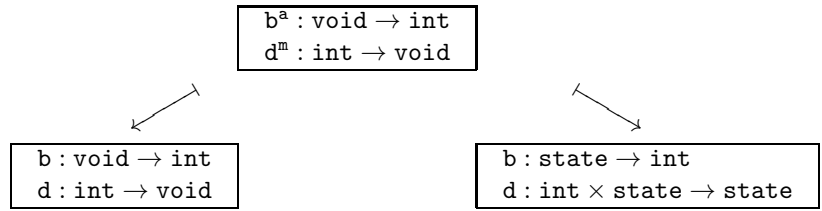
  \begin{array}{l}
  \texttt{b}^\texttt{a}:\texttt{void}\to\texttt{int} \cr
  \texttt{d}^\texttt{m}:\texttt{int}\to\texttt{void} 
  \end{array} }
  & \\
\cline{2-2}
  \xymatrix@R=1pc{&& \ar@{|->}[dl] \\ && \\ } 
  && 
  \xymatrix@R=1pc{\ar@{|->}[dr] && \\ && \\ }\\
  \cline{1-1} \cline{3-3}
  \multicolumn{1}{|c|}{
  \begin{array}{l}
  \texttt{b}:\texttt{void}\to\texttt{int} \cr
   \texttt{d}:\texttt{int}\to\texttt{void} 
  \end{array} }
  && 
  \multicolumn{1}{|c|}{
  \begin{array}{l}
  \texttt{b}:\texttt{state}\to\texttt{int} \cr
   \texttt{d}:\texttt{int}\times\texttt{state}\to\texttt{state}
  \end{array} } \\ 
\cline{1-1} \cline{3-3}
  \end{array} $$

  \vspace{1cm}
  
\section*{CONCLUSION}

\mypar{We propose an abstract algebraic framework for logic.}

  \begin{itemize}
  \item A \emph{simple} framework:
  \begin{itemize}
  \item A diagrammatic \emph{logic}
    is a \emph{sketch}.
  \item A diagrammatic \emph{logical rule} 
   is a \emph{fraction}. 
  \end{itemize}
  \item A \emph{homogeneous} framework:
    \\ $\quad$ ``the logic of logics is a logic''.
  \item A \emph{category} of logics:
    \\ $\quad$ morphisms of logics are fractions of sketches. 
  \end{itemize}


\end{document}